# An Effective Parallel Program Debugging Approach Based on Timing Annotation

Yun Chang
Dept. of Computer Science
National Tsing-Hua Univeristy
Hshinchu City, Taiwan
takojoyce@gmail.com

Hsin-I Wu
Dept. of Computer Science
National Tsing-Hua Univeristy
Hshinchu City, Taiwan
line 5: email address

Ren-Song Tsay
Dept. of Computer Science
National Tsing-Hua Univeristy
Hshinchu City, Taiwan
line 5: email address

*Abstract*—We propose an effective parallel program debugging approach based on the *timing annotation* technique. With prevalent multi-core platforms, parallel programming is required to fully utilize the computing power. However, the non-determinism property and the associated concurrency bugs are notorious and remain to be great challenge to designers. We hence propose an effective program debugging approach using the timing annotation technique derived from the deterministic Multi-Core Instruction Set Simulation (MCISS) technology. We hence construct a deterministic execution environment for parallel program debugging and devise a few unique, effective and easy-to-use parallel debugging functions. We modify QEMU and GDB to implement and demonstrate our proposed idea. The usage of our debugger is almost identical to the conventional GDB debugger. Therefore, users may learn how to use the tool seamlessly.

## I. Introduction

As multi-core systems are becoming ubiquitous, parallel programming is a critical technology to fully utilize the computing potential for ever more performance hungry tasks. However, parallel programming is notorious for its non-determinism nature [1] and the associated concurrency bugs, if exist [2]. Proper parallel program implementation is indeed challenging, whereas perplexing concurrency bug debugging is even more challenging.

Different from sequential program executions, parallel program executions are non-deterministic mainly due to CPU execution speed variations and the pseudo-randomness of the system OS scheduling. Even with the same input, different execution runs of a parallel program may produce different results. Essentially, this is because the uncontrollable interleavings of execution threads of parallel programs result in different shared-data access order. The non-deterministic shared-data access sequence, not following the designer's original intention, is the primary source of concurrency bugs for which the most difficult challenge is that the bugs are not reproducible in every execution.

The non-determinism property not only perplexes programmers on designing correct parallel programs but also incapacitates the traditional cyclic debuggers which have been widely and successfully used for sequential program debugging [3]. When encountering a bug in a sequential program, programmers can iteratively execute the program for debugging; however, when a concurrency bug occurs in parallel program execution, the designer may not even be able to reproduce the bug because of no control of the thread interleavings and the shared-data access order.

Nevertheless, we learned from system simulation technologies such as MCISS (Multi-Core Instruction Set Simulation) technology [4][5][6][7][8] which employs a successful timing annotation technique and can guarantee repeatable execution results. Essentially, this timing annotation technique explicitly computes the execution time of each running thread using the timing models of involved computing elements. With deterministic timing models, the execution time computed is also deterministic and is used to determine the chronological order of shared data accesses and synchronization events. Note that a synchronization event is an event whose execution order may affect the overall shared-data access order. To ease later discussions, we simply assume only shared data accesses need to be synchronized. In fact, event synchronizations in principle are also working on top of some event-specific shared variables.

Therefore, equipped with the timing annotation method, we can build a deterministic execution environment for parallel program debugging. Along with the program execution, the collective execution time is calculated and annotated to the shared data access point. The annotated time is then used to determine the data access execution sequence which in turns also deterministically regulates the thread interleaving order. We implement our proposed approach on the widely adopted emulator QEMU [9] by incorporating the timing annotation feature.

Practically, the timing annotation technique can facilitate many unique and useful debugging functions for parallel programs in addition to just constructing a deterministic execution environment. In general, the traditional sequential debugger cannot debug parallel programs properly because of lacking the notion and control mechanism of interleavings. However, with the timing annotation technique, one can implement effective parallel program debugging functions by monitoring and managing not only the variable (data) values but also the associated annotated time values. This fact of being able to control the value and time of each variable in debugger implies that we can easily observe the interleaving order or even modify the order to trigger certain desired effects.

For implementation, we simply enhance the existing popular open-sourced GDB [3] for our purpose. With the timing-annotated deterministic execution environment and the enhanced effective parallel-program debugging functions, our approach offers a user-friendly debugger seamlessly integrated with the conventional sequential debugger.

The remainder of this paper is organized as follows. Section 2 reviews related work. Section 3 introduces the methodology of timing annotation. Section 4 and 5 present the implementation of building a deterministic execution environment and designing debugging functions with timing annotation. Section 6 shows the implementation and testing. Section 7 discusses a few real cases to demonstrate the usage of our proposed tool, followed by a conclusion in Section 8.

## II. Related Work

For parallel program debugging, the main task is to reproduce the buggy execution. When the buggy execution can be re-produced, users are able to start the cyclic debugging process. We briefly review two main approaches: record-replay [11][12][13][14][15] and reverse execution [16][17] in the following.

In order to reproduce the buggy execution, some believe that reproducing the bug occurring condition and environment is the most direct way [11][12][13][14][15][16][17]. By reconstructing the buggy execution, users then can debug the program repeatedly. Raw information generated from previous executions is used to reconstruct the buggy execution. With the provided information, these approaches reproduce the buggy execution for users to perform the debugging process.

For record-replay approach [11][12][13][14][15], the approach first records the execution trace, and then replays the recorded execution. The idea is to perform debugging right after encountering bugs. By recording the execution's runtime states, such as variables' values, registers, and memory content, the record-replay approach can reproduce the original execution precisely. However, to reproduce the execution the detailed recording requires huge log file size. Also, a designer will need to first identify the concurrency bug first, which may not be trivial by itself.

For reverse execution [16][17], the approach utilizes the core dump file and other fragments generated from the failed execution to reconstruct the buggy execution. When the execution of a program failed, the system will generate a core dump file for users to inspect the execution's status. With the core dump file and the back-tracing algorithm, the approach can reconstruct the buggy execution. Nevertheless, the approach heavily relies on the designed back-tracing algorithm to guarantee the correctness of the reconstructed execution.

## III. Timing Annotation technique

As previously discussed, the real systems normally do not allow user intervention to control the parallel executions, and therefore the non-determinism property greatly causes challenges for parallel program debugging. Nevertheless, we learned from some effective simulation technologies [4][5][6][7][8] that the timing annotation technique can provide a deterministic parallel execution environment and facilitate effective parallel program debugging.

The idea of timing annotation is to compute the execution time of each executing component based on deterministic timing models and tag the collective execution time to the synchronization points of interest. With the annotated timing value on each synchronization point, one may easily determine the execution order (or interleaving order) of the synchronization point in a simulator. With deterministic timing models and calculation, the execution order is deterministic and hence the parallel execution is guaranteed to be deterministic.

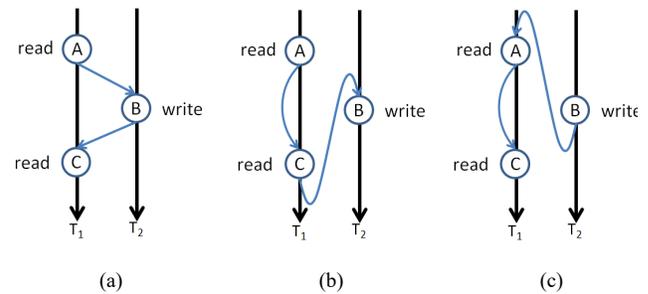

Fig. 1. An illustartion of the same code section with different interleavings. A, B, and C represent access points of a certain shared variable. The non-deterministic executions result in three possible interleaving orders as shown in (a), (b) and (c).

We use the example in Fig. 1 to illustrate how timing annotation works in a parallel program simulation. Suppose that the interleaving order, A(read) → B(write) → C(read) as shown in Fig. 1(a), is the designer's intended execution order, where A, B, and C are access points of a certain shared variable. Practically, the non-determinism property of parallel program may cause unexpected orders to be executed randomly in different executions, as shown in Fig. 1(b) and 1(c). With timing annotation, we know the time for synchronization point A to be executed is at 100 time unit point and the time synchronization point B to be executed is at 120 time unit point. Therefore, A is to be executed first, and then the timing annotation method determines to tag C of value 123. Since the timing annotated value of each synchronization point is always the same even at different execution run if we have the same execution starting time and input, the order is always deterministically in the A(read) → B(write) → C(read) order as shown in Fig. 2.

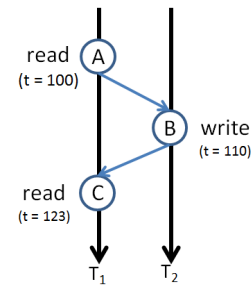

Fig. 2. An illustartion the same example of Fig. 1 with timing annotated value on each access (synchronization) point.

With timing annotation, we not only can perform deterministic parallel program executions but also can offer many unique debugging features for parallel programming. With the annotated information of time, an enhanced debugger from existing debugging tools, such as GDB, can allow users explicitly identify or even control the execution flow. When needed, a user can manually overwrite the timing annotated values of certain synchronization points and force triggering suspicious bugs without need to refactor the original source code frequently.

In the following sections, we will first discuss how to build a deterministic execution environment with timing annotation, followed by how to use that environment and timing annotation to design the debugging functions for parallel program debugging.

## IV. Building a Deterministic Environment

As discussed before, with deterministic timing models and timing annotation on synchronization points, we may easily determine the chronological order of all synchronization points. With the chronological order determined, executions are deterministic and always produce the same result with the same input. Next, we elaborate the three steps to build a timing-annotation-based deterministic execution environment: (1) identify sync points, (2) perform timing annotation, and (3) perform synchronization.

### A. Identify Sync Points

First, we need to find where in the program to perform synchronization. For simulation technology, a sync point is referred to a specific code point whose execution order may affect the results of execution. In our work, a sync point is a shared data access point in a thread whose execution order may affect the execution results of other threads and the final results.

For example, global variables can be accessed by any threads and are used to exchange data between threads; thus, each global variable access point has to be regarded as a sync point.

### B. Perform Timing Annotation

With sync points identified, we need to know how to perform synchronization next. Essentially, we use the timing-annotated value of each sync point to determine the chronological execution order of each sync point and hence the interleaving order of all threads.

For implementation, we leverage the MCISS (Multi-Core Instruction Set Simulation) [4][5][6][7][8], which is a mature multi-core simulation technology. The MCISS executes programs deterministically by synchronizing on each sync point with the annotated time value, which is calculated from timing models. A timing model essentially describes how the execution time is spent on each execution component along the execution path between two consecutive sync points. By accumulating the execution time along the execution path, the timing annotation step basically annotates the collective execution time value to each sync point. With the annotated timing value on each sync point, we then deterministically determine the chronological order and hence the execution order of all sync points.

In practice, one may adopt timing models of different timing granularities in various precision; however, no matter what timing model is selected, once the model used is deterministic, the timing annotated values are deterministic. In fact, the choice of timing model is not critical for debugging purpose. Although the calculated execution time may be off slightly, practically the same program running on different hardware platforms may also have different execution times. Therefore, how to create a deterministic execution environment is most critical for debugging purpose practically.

### C. Perform Synchronization

With sync points identified and timing annotation determined, one can then determine the chronological order of sync points for execution. There are a few known synchronization techniques [5][6][8], and we adopt the collaborative approach in our implementation. For the collaborative synchronization approach, whenever a thread hits a sync point, the thread relinquishes its execution and queues the execution to a central scheduler. Then the central scheduler will examine from the queued threads and determine which thread with the smallest timing annotation value to continue the execution. For details, one may refer to the MCISS papers [4][5][6][7][8].

With the three steps outlined above, one can construct a deterministic execution environment. Next, we will discuss how to use the deterministic environment and the timing annotation technique to design parallel program debugging functions.

## V. Designing Debugging Functions

Based on the deterministic execution environment discussed above, we devise a few new debugging functions specific for parallel programs while adopting conventional sequential program (or cyclic) debugging functions. The key difference to the conventional sequential program debugger is that our approach can monitor and change not only variable values but also timing annotated values of synchronization points.

### A. Adapt Traditional Cyclic Debugging Functions

Traditional cyclic debugging techniques work well in debugging sequential programs, but these techniques are incapable of reproducing the concurrency bugs in parallel programs due to no notion and control of interleavings. However, the timing-annotation-based deterministic environment essentially fixes the interleaving order so that the parallel program executes in a way similar to how a sequential program executes. Next, we elaborate how we adapt the three frequently used debugging functions in GDB [3], including breakpoint, step, and runtime state modification, to the deterministic parallel program execution environment.

*1) Breakpoint*

During the debugging process, breakpoint allows users to stop the execution at any intended point in a program. When the execution stops at breakpoint, users can then inspect related information of the particular spot, such as the value of program counter, registers, memory, etc. In sequential program debugging, users often leverage full control of the execution flow with breakpoints. The execution will stop at the specified breakpoints and allow users to examine the execution contents and to identify possible bug causes.

Nevertheless, setting a breakpoint in a parallel program will be chaos if the debuggers have no notion of concurrency. For example, as shown in Fig. 3(a), users set a breakpoint $\alpha$ in the thread $T_b$. When the thread $T_b$ breaks at $\alpha$, users will have no idea about where the thread $T_a$ may pause at and what variable values are affected. The reason is that the traditional GDB is ignorant of the concurrent executions; hence, other threads without breakpoints will continuously execute until GDB sends a signal to stop them.

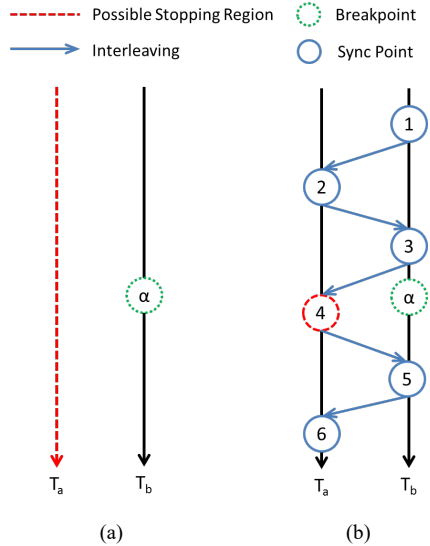

Fig. 3. An illustratation of setting a breakpoint (a) in a non-deterministic execution environment, and (b) in our proposed deterministic execution environment.

Now, we explain how our deterministic approach resolves the concurrency ignorance issue. As illustrated in Fig. 3(b), with the breakpoint $\alpha$ set at the same spot as in Fig. 3(a), if the thread $T_b$ is set to break at $\alpha$, then users will find the thread $T_a$ has to stop at sync point 4 since $\alpha$ is located in a local code segment between the sync points 3 and 5, and the interleaving order is 1 → 2 → 3 → 4 → 5 → 6. With the sync point 4 reached and $T_a$ stopped, $T_b$ resumes its execution from the sync point 3. Note that according to the simulation rule, $T_a$ will remain stopped at the sync point 4 until $T_b$'s execution hits the sync point 5. Because the interleaving is fixed by the chronological order, our approach guarantees that users know where other threads shall stop when a breakpoint is set in a particular thread.

*2) Step Function*

In a conventional debugger, the step function allows users to track the execution following the source code line by line. By stepping through a program, users can promptly inspect the program's status. However, the step function cannot be applied to debugging parallel program executions directly. Similar to the same interleaving controllability issue of the aforementioned break function, we will need to apply the timing annotation and synchronization techniques for proper management.

We use the example in Fig. 3(a) to illustrate the issue. Suppose that a user uses step function after the thread $T_b$ hits the breakpoint $\alpha$. If with no control of interleaving, when the user is managing advancing one execution step in $T_b$, the thread $T_a$ may have executed many lines of code and creates unexpected interleaving order.

We now use the example in Fig. 4 to illustrate how we modify the step function for parallel debugging. With the fixed interleaving order, when a user wants to perform step executions after the breakpoint $\alpha$, the thread $T_a$ shall stop at the sync point 4, whose annotated timing value is in between the values of the sync points 3 and 5 and the breakpoint $\alpha$ is located in between the execution path from the sync point 3 to the sync point 5. When the thread $T_b$ hits but not yet executes the sync point 5, the stepping execution should switch to $T_a$ because the chronological order of the sync point 4 is before that of the sync point 5.

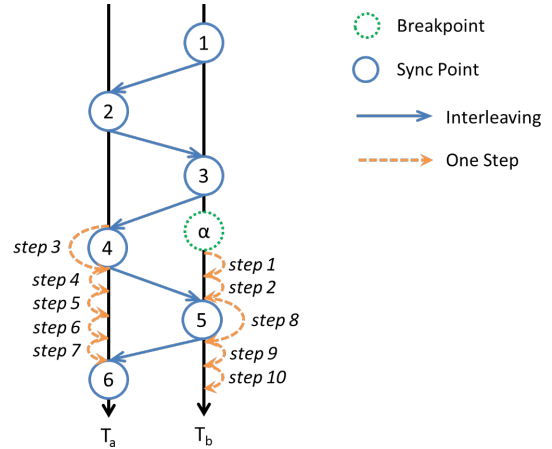

Fig. 4. An illustratation of performing step function in our proposed deterministic execution environment with the stepping sequence listed.

In fact, in our deterministic execution environment, the stepping executions for parallel programs become very similar to the stepping executions in sequential programs. The benefit of this approach makes the debugging of parallel programs become much easier.

*3) Runtime State Modification*

In the conventional cyclic debuggers, users are allowed to modify the values of variables, a part of the runtime process states. With process states modified, the execution may lead to a new path and the new execution path may trigger certain behaviors and help clarifying possible bug causes.

In contrast, the runtime state modification although can still be performed for parallel program debugging under the conventional cyclic debuggers, the results may not be reproducible or expected due to the non-determinism nature as discussed previously.

However, with the timing annotation technique and the resulted deterministic execution environment, executions of the same program with the same input always follow a unique interleaving order and always produce the same result. Therefore, with same modifications at same spots, our environment guarantees determinism and always produces consistent results.

Note that in our approach, our proposed parallel program debugger can also allow users to modify the annotated timing values of sync points and directly change the interleaving order. This is one of the most notable features of our approach as compared to the limited variable value modification function in the conventional cyclic debuggers. Details are to be discussed later.

### B. New Features for the Proposed Debugger

By providing the deterministic execution environment, our approach greatly simplifies parallel program debugging tasks. In the following, we introduce a few unique and effective parallel program debugging features which are made possible under our proposed deterministic approach. The first one is the sync-step function and the second one is the runtime interleaving modification feature.

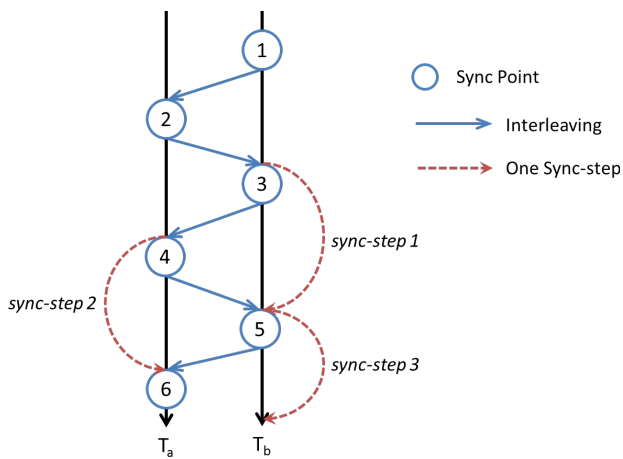

Fig. 5. An illustration of performing sync-step function in our proposed deterministic execution environment with the stepping sequence listed.

#### 1) Sync-Step Function

In contrast to the step function shown in Fig. 4, the sync-step function is designed for users to quickly proceed the execution of one thread to the nearest next sync point without need to frequently perform 'step' commands, while the other threads remain stopped. Once the thread hits the sync point, users can switch to other threads to perform other debugging operations. Note that for the sync-step execution, all sync points will follow the interleaving order determined by the timing annotated values. For example, in Fig. 5, when the thread $T_b$ encounters the sync point 5 after executing sync-step from the sync point 3, the execution shall switch to the thread $T_a$ and execute from the sync point 4.

#### 2) Runtime Interleaving Modification

With the fixed interleaving order, our approach does produce a deterministic parallel program execution environment. Then one side effect is that if users want to trigger a different interleaving order then traditionally the only way is to try different inputs or to modify source code. Now, equipped with the timing annotation feature, our proposed debugger allows users to modify the timing annotated values of sync points and therefore change the chronological order of sync points during the debugging process.

We use the example in Fig. 6 to demonstrate the process of changing interleaving order by modifying the timing annotated values. Suppose that the original interleaving order is $1 \rightarrow 2 \rightarrow 3 \rightarrow 4 \rightarrow 5 \rightarrow 6$, with each sync point attached a timing annotated value as shown in Fig. 6(a). Assume that the user suspects a bug may occur if the execution order of the sync point 2 to 3 is not right, then he can break at the sync point 2, and modify the timing annotated value of the sync point 2 from a value smaller than that of the sync point 3 to a value which is larger. Then as shown in Fig. 6(b), the interleaving order becomes $1 \rightarrow 3 \rightarrow 2 \rightarrow 4 \rightarrow 5 \rightarrow 6$ after changing the timing annotated value of the sync point 2 from 25 to 30. Note that since the execution order is changed, the timing annotated values of the sync points 4 and 6, which follow the sync point 2, will need to be updated accordingly. Importantly, a user may repeat this new execution order and produce consistent results if same modification is repeatedly done.

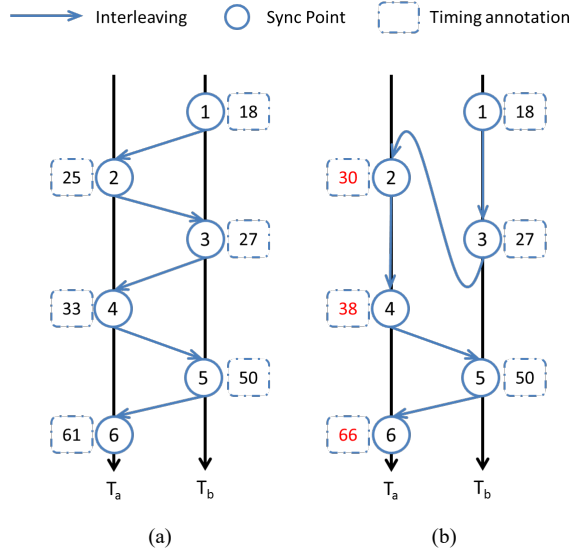

Fig. 6. An illustration of modifying the interleaving from (a) to (b) by changing the timing annotation of sync point 2 to 30. The timing annotation of sync point 4 and 6 also changed after the modification.

## VI. IMPLEMENTATION AND TESTING

In this section, we present the implementation and testing result as a demonstration of our proposed timing-annotation-based debugging framework. First, we show how we extend QEMU [9] to a deterministic MCISS (Multi-Core Instruction Set Simulation) for parallel program debugging. Then, we verify the determinism of the modified QEMU.

### A. Modify QEMU

We incorporate the timing annotation technology into QEMU 2.9.1's user-mode, by setting the target machine to be ARMv7 RISC ISA. Basically, QEMU is originally a full-featured emulator which supports multiple architectures. In order to guarantee determinism, we modify QEMU by adding timing models for calculating timing annotations, and synchronization mechanism.

For sync points, we set all global variable access points as sync points. We identify these variables from the address space, and we perform synchronization before every load/store function in the assembly code level to detect global data accesses. For debugging, we integrate the remote debugging tool GDB-Stub [10] for ARM architecture, which is built in QEMU, as gdbserver, and GDB for ARM in Linux as gdbclient.

### B. Testing Setup and Results

We use a host machine equipped with eight-core Intel Xeon E5-2650 CPUs running at 2.0 GHz and Ubuntu Linux14.04 OS as the testing environment. To verify determinism, we use standard RACEY [18] test case.

#### 1) Detetminism

We verify our environment's determinism on RACEY, which generates a signature 32-bit value that is sensitive to the order of shared-data accesses. With 100 separated execution run of RACEY, our environment always produces the same signature value.

In contrast, the original QEMU, with no timing model enhancement, produces various signatures while executing RACEY.

## VII. CASE DEMONSTRATION

## VIII. CONCLUSION

In this paper, we propose an effective parallel program debugging approach based on the timing annotation technique. Using timing annotation, we fix the interleavings of parallel programs, and thus guarantee determinism. With our deterministic execution environment for parallel program, we adapt the conventional cyclic debugging features to be fully-utilized for parallel program debugging. Moreover, we devise a few new functions that are specific for debugging parallel program. We implement our approach by modifying QEMU and GDB and show the testing result in the end of the paper.